# Power-law Signatures and Patchiness in Genechip Oligonucleotide Microarrays

Radhakrishnan Nagarajan

University of Arkansas for Medical Sciences, 4301 W. Markham, Little Rock, AR 72205, USA. Email: nagarajanradhakrish@uams.edu

**Abstract.** Genechip oligonucleotide microarrays have been used widely for transcriptional profiling of a large number of genes in a given paradigm. Gene expression estimation precedes biological inference and is given as a complex combination of atomic entities on the array called probes. These probe intensities are further classified into perfect-match (PM) and mismatch (MM) probes. While former is a measure of specific binding, the latter is a measure of non-specific binding. The behavior of the MM probes has especially proven to be elusive. The present study investigates qualitative similarities in the distributional signatures and local correlation structures/patchiness between the PM and MM probe intensities. These qualitative similarities are established on publicly available microarrays generated across laboratories investigating the same paradigm. Persistence of these similarities across raw as well as background subtracted probe intensities is also investigated. The results presented raise fundamental concerns in interpreting Genechip oligonucleotide microarray data.

## 1 Introduction

Oligonucleotide Genechip microarrays [1, 35, 36] have been used widely for transcriptional profiling of large number of genes across distinct biological paradigms including (i) *stem cell differentiation* [27, 47], (ii) *molecular portraits and heterogeneity in tumors* [43, 50], (iii) *Aging and neurobiology* [13], (iv) *infectious disease research and environmental applications* [31]. Prevalence of such high throughput assays can espe-cially be attributed to the rapid sequencing of genomes [11]. A recent multiple-laboratory and multi-platform study [26] established the superiority of oligonucleotide microarrays from accuracy and precision standpoints. Unlike classical biological approaches, microarrays can be used to model functional relationships between genes, hence provide *system-level* understanding [30] of the paradigm [14, 59]. There is also the possibility of oligonucleotide arrays being used as active screening tools in clinical settings in the near future [21].

Developing suitable computational techniques for meaningful interpretation of oligonucleotide gene expression data is one of the major challenges and precedes biological inference. Gene expression is estimated as a complex combination of atomic entities on the array called *probes* [45]. While several a1gorithms have been proposed for gene expression estimation and subsequent higher level analysis [2, 3, 24-26, 34, 46, 48], understanding the qualitative behavior at the probe level is still *incomplete*. Probes are broadly classified into *perfect match* (PM) and *mismatch* (MM). The former is a measure of *specific binding* whereas the latter is a measure of *non-specific binding* and used as an internal control (Sect. 1.1) [1, 35, 36]. While PM and MM probes are biologically distinct by very design they are spatially proximal on the array. Several statistical techniques have been proposed for gene expression estimation and subsequent higher-level analysis. While some techniques use perfect as well as mismatch probes [2, 3, 34], others have encouraged using the perfect match probes only [24, 25] in the estimation procedure. The choice of the latter was possibly inspired by [38], which pointed out that arithmetic subtraction of (PM, MM) probe intensities may not translate into biological subtraction. The qualitative behavior of the MM probes has especially proven to be elusive.

The objective of the present study is to investigate qualitative similarities in the distributional signatures and local correlation structure across the perfect-match and mismatch probe intensities. Qualitative similarities are demonstrated on the raw as well background subtracted (PM, MM) probe intensities in publicly available Genechip arrays generated across laboratories investigating the same biological paradigm [26]. These qualitative similarities to our knowledge have never been reported and raise fundamental concerns in interpreting oligonucleotide gene expression data and higher level analyses such as (*a*) gene expression estimation and normalization [2, 3, 6, 24, 25, 34, 46, 48, 58]. (*b*) inferring functional relationships and network structure [14, 59] (*c*) ontology [5] and (*d*) expression quantitative trait loci (eQTL) [28] The present study is especially encouraged by our (*i*) recent research on various aspects of microarray gene expression analysis [39, 40] and growing evidence of (*ii*) hybridization interactions/multiple targeting of the probes [42, 57, 60]; (*iii*) spatial artifacts [52] and (*iv*) redefinition of probe-transcript relationship [16, 33] in oligonucleotide Genechip arrays .

The chapter is organized as follows. In Sect. 1.1, a brief introduction to Genechip oligonucleotide microarrays along with the associated terminologies is provided. Qualitative similarities along with power-law and exponential approximations to the PM and MM probe intensity distributions is investigated in Sec. 2. Qualitative similarities in local correlations/patchiness across PM and MM probe intensity matrices is investigated in Sec. 3. The choice of multiscale decomposition for accomplishing the same is also explored. The impact of the findings in the present study on gene expression estimation and subsequent higher level analyses is discussed in Sect. 4.

## 1.1 Oligonucleotide Genechip microarrays

Oligonucleotide Genechip microarray [1, 35, 36] comprise of a large number of atomic entities called *probes* [45] arranged as a rectangular matrix. Each probe is an *oligomer*, i.e. around ~25 nucleotides long, (e.g. 5'-GTGATCGTTTACTTCGGTGCCACCT-3'). A set of (~16 to 20) probes also called a *probeset*, represents a particular *transcript* on the array. The term transcript is generic and can represent either a *gene* or an *expressed sequence tag (EST)*. Probes can be broadly classified into *perfect-match* (PM) and *mismatch* (MM) probes. PM probes correspond to a short region of the transcript and are designed to be complementary to the *target sequence* [1, 35, 36], hence ideally a measure of *specific binding*. The nucleotide content of an MM probe is the same as that of the corresponding PM probe except for the middle most nucleotide, which is changed deliberately. Thus MM probes are used as an internal control to *assess non-specific binding*. Gene expression $g^t$ of a transcript $t$ on the array is given as a complex combination of the corresponding (PM and MM) or PM only probe intensities [2, 3, 24, 25, 34, 48]. An example of PM, MM and their target probe is shown below for clarity.

**Example** PM, MM and target probe:

```
PM      (5' G T G A T C G T T T A C T [T] C G G T G C C A C C T 3')
MM      (5' G T G A T C G T T T A C T [C] C G G T G C C A C C T 3')
Target  (5' C A C T A G C A A A T G A  A  G C C A C G G T G G A 3')
```

Analysis of oligonucleotide microarrays begins by extracting the raw (PM, MM) probe intensities from the .CEL files [1] (Affymetrix Technical

Manual, Santa Clara, CA). Subsequently, these are *background subtracted* [2, 3, 24-26, 34, 48, 58] to minimize contributions of non-biological factors to the probe intensity/gene expression. In the present study, we investigate the qualitative similarities of the raw as well as background subtracted [3, 24, 25] (PM, MM) probe intensities. Such an approach is useful in rejecting the claim that the observed qualitative similarities are an outcome of not subtracting the background. Background subtraction is accomplished with Bioconductor [17] implementation of two popular algorithms namely: MAS 5.0 [3] and RMA [24, 25]. Consider the PM $\pi^{pmt} : \pi_1^{pmt} ... \pi_{20}^{pmt}$ and MM $\pi^{mmt} : \pi_1^{mmt} ... \pi_{20}^{mmt}$ probe intensities corresponding to a transcript *t*. The gene expression of that transcript is a mapping of $\pi^{pmt}$ and $\pi^{mmt}$ onto a single value $(g^t)$ by a chosen estimation procedure *f*, represented by $(\pi^{pmt}, \pi^{mmt}) \xrightarrow{f} g^t$.

It is important to note that depending on the choice of the estimation procedure *f*, gene expression $(g^t)$ is either a *linear* or *nonlinear* combination of $(\pi^{pmt}, \pi^{mmt}, t = 1...20)$. An example of linear and a nonlinear estimation procedures assuming two $(\pi^{pmt}, \pi^{mmt})$ probes per transcript $(g^t)$ and their impact on the distributions is shown below for clarity.

**Example** Mapping from probe intensity to gene expression

*(a) Linear estimation procedure*

$$f: g^t = (2\pi_1^{pmt} + 3\pi_2^{pmt}) - (0.5\pi_1^{mmt} + \pi_2^{mmt})$$

In (a), gene expression estimation is given as a difference of the corresponding PM and MM intensities. If $(\pi^{pmt}, \pi^{mmt})$ are normally distributed then $(g^t)$ is *normally* distributed.

*(b) Nonlinear estimation procedure*

$$f: g^1 = (2\pi_1^{pm1} + 3\pi_2^{pm1})^2 - (0.5\pi_1^{mm1} + \pi_2^{mm1})^2$$

In (b), gene expression estimation is given as a difference of the square of cor-PM and MM intensities. Unlike (a), even if $(\pi^{pmt}, \pi^{mmt})$ are normally distributed $(g^t)$ is *not normally* distributed in (b).

***Remark 1*** *From the above section we note the following important points:*

*(i)* *Gene expression is estimated as a complex combination of (PM and MM or PM only) probe intensities using an estimation procedure f. Thus conclusions drawn about the statistical properties such as distributional profiles at the gene expression level are dependent on the as-* *sumptions behind that particular estimation procedure f. However, conclusions drawn at the probe intensity level is independent of the estimation procedure f.*

*(ii)* *PM and MM probe intensities although biologically distinct are located physically adjacent to each other on the array (i.e. spatially proximal).*

*(iii)* *Spatial information preserved at the probe intensity level is lost at the gene expression level. Since one of the objectives of the present study is to understand the qualitative similarities in local non-random structure/patchiness across the array, retaining the spatial information is crucial.*

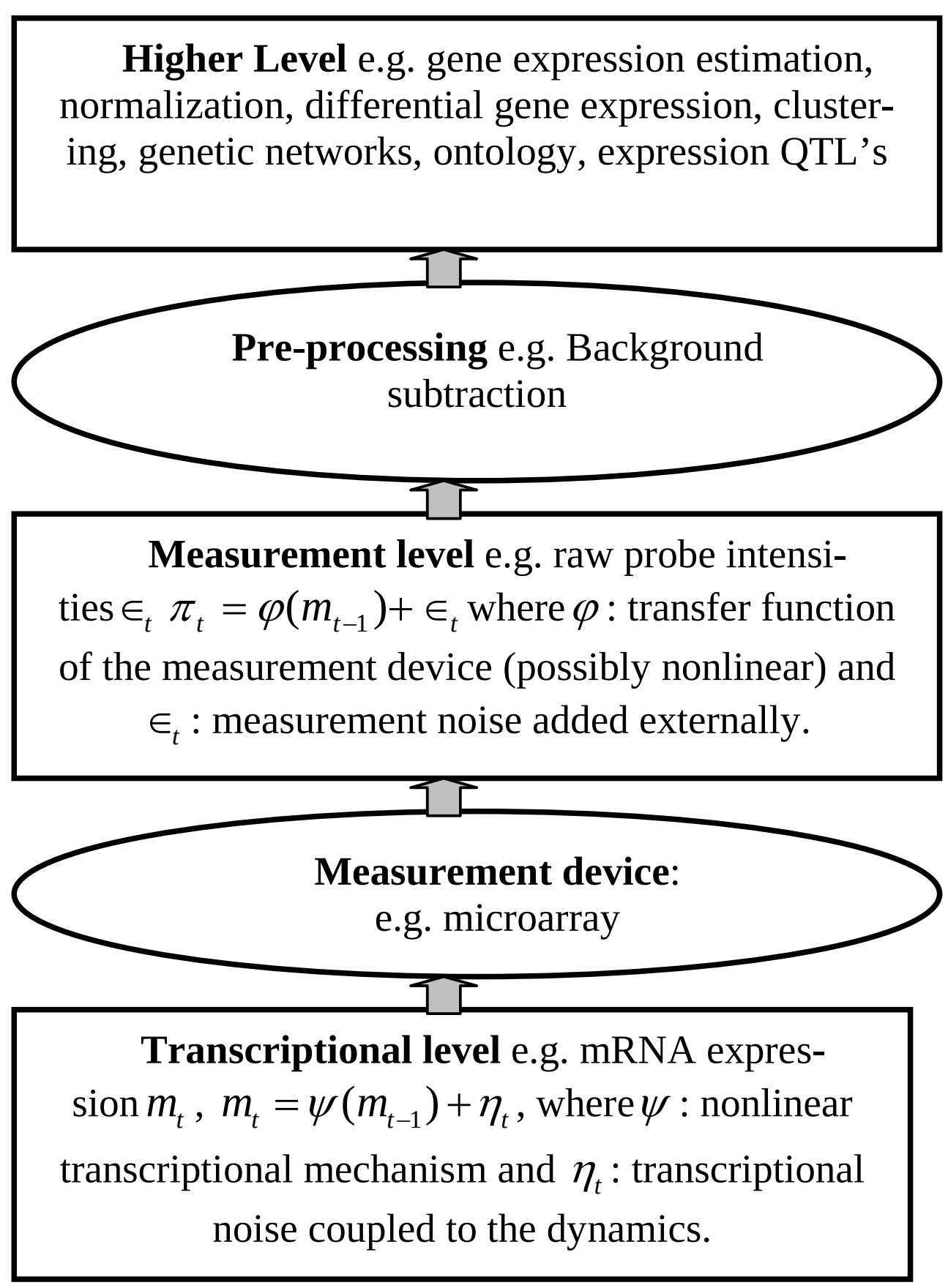

**Fig.1.** Schematic diagram representing the contribution of various factors to probe intensity and gene expression estimates.

A significant number of studies [22, 32, 56] have argued in favor of power-law distributional approximation to microarray gene expression data and attributed the same to biological factors governing gene expression. The authors in [22] demonstrated power-law (pareto-like) distribution in gene expression across genomes. They attributed such a behavior to common probabilistic mechanism in the gene expression process conserved in eukaryotic evolution. The authors in [32] claimed that gene expression distributions across several microarray platforms show close similarities to power-law behavior. Their findings also claimed that the variance of the log spot intensities were proportional to the genome size. In [56], the authors demonstrated persistence of power-law signatures in microarray gene expression from bacteria (Escherichia Coli) to humans (Homo Sapiens) across distinct biological conditions. Such a behavior was attributed to universality in transcriptional organization across genomes.

In this section, we explore biological and non-biological factors that contribute to the distribution of probe intensities; hence gene expression estimates (Remark 1). A schematic diagram representing the microarray data acquisition process and subsequent higher level analysis is shown in Fig. 1 [41]. Specific details such as array layout, probe descriptions, hybridization protocols, laser scanning and image segmentation are intentionally excluded in Fig. 1 and can be found elsewhere [1, 35, 36]. Oligonucleotide microarrays can be regarded as *measurement devices* or transducers that map the true transcriptional activity (i.e. mRNA expression) onto a measurement value (i.e. raw probe intensities). The data acquisition (Fig. 1) is accompanied by considerable noisiness $(\eta_t, \in_t)$ and nonlinearities $(\varphi, \psi)$ at the transcriptional and the measurement levels, Fig. 1. Transcriptional noise is coupled to the dynamics of the system, hence *biological*. It can be attributed to uncertainty in gene expression [12, 29, 53]. However, measurement noise is uncoupled to the dynamics of the biological system, hence *non-biological*. Biological systems by their very nature are nonlinear feedback systems [15, 18, 51]. An example of nonlinearity ($\psi$) in the case of gene expression is that of transcriptional cooperativity [15, 18], where promoters work in tandem to facilitate transcription. The actual mRNA expression and those output by a measurement device such as an oligonucleotide microarray need not necessarily be linearly related. The measurement device is often accompanied by an associated transfer function ($\varphi$) possibly nonlinear, that maps the true biological activity (i.e. mRNA activity) onto the raw (PM, MM) probe intensities. It is important to appreciate the fact that ($\psi$) is *biological* whereas ($\varphi$) is *non-biological*.

***Remark 2*** *From the above section we note the following important points:*

*(i)* *Biological as well as non-biological factors can contribute to the probe intensity/gene expression estimates, Fig. 1.*

*(ii)* *The distribution at the probe intensity is governed by the*

*(a) distribution of the transcriptional and measurement noise* $(\eta_t, \in_t)$ *which can be Gaussian (i.e. additive process) or non-Gaussian (e.g. multiplicative process)*

*and*

*(b) nonlinearities at the transcriptional and measurement levels* $(\psi, \varphi)$.

*Therefore, even if the true biological process (i.e. mRNA levels) is normally distributed, the distribution of the measured probe intensities (PM, MM) is likely to be skewed. The skew in the distribution of the probe intensities is also accentuated by their non-uniform nucleotide content which in turn governs the binding efficiencies, hence their expression [57, 58, 60]. Artifacts due to non-specific binding [16, 33, 42] and spatial gradients [52] also contribute to the probe intensity/gene expression estimates.*

***Remark 3*** *While the distribution of the raw probe intensities are governed by the factors listed under Remark 2, those of gene expression has significant contribution from the factors under Remark 2 as well as the estimation procedure f (Remark 1). Therefore, the qualitative properties at the gene expression level need not reflect those at the probe intensity level.*

**Data** The microarrays considered in the present study are publicly available and were generated in a recent study [26] (Affymetrix, Human Genome U133 set, i.e. HGU133A, 22283 transcripts) on comparing gene expression results across microarray platforms and laboratories. The corresponding .CEL files [1] containing the PM and MM probe intensities is in the form of a rectangular matrix with dimensions 356 x 712. All entries in this matrix which had zero intensity were forced with uncorrelated random numbers in order to reject any spurious correlation. Considering replicate arrays across laboratories rejects the claim that the observed results are not an outcome of experimental protocols adopted by a particular laboratory.

## 2. Power-law distributional approximations to PM and MM probe intensities

Array-wide gene expression has been widely reported to exhibit a significant skew towards lower expression values and a decaying trend with increasing magnitude of expression. Several parametric distributions can be used to model such a decaying trend [9]. Static nonlinear transforms such as Box-Cox normality transforms $\xi(x) = (x^{\lambda} - 1)/\lambda$ [8] have been used widely in statistical literature to argue in favor of near-normality assumptions. The log-transform in conjunction with 2-fold cut-off used widely in microarray community for identifying differential gene expression is the limiting case of classical Box-Cox normality transforms, i.e. $\lim_{\lambda \to 0} \xi(x)$.

This in turn implicitly assumes log-normal distribution of the gene expression values. Two popular distributions used widely to model decaying trends include the exponential and power-law distributions. The parameters of both the distributions can be attuned so as to capture the decaying trend with increasing magnitude. However, these two classes of distributions have marked differences in their statistical properties. Unlike exponential distribution, the power-law distributions exhibit *scale-invariance*, where the basic shape of the distribution does not alter with scaling. Let $p(k) \sim k^{-\gamma}$ then we have $p(\theta.k) \sim \theta^{-\gamma}.k^{-\gamma} = \theta^{-\gamma} p(k)$ i.e. the distribution of $p(k)$ resembles that of $p(\theta.k)$ other than for a constant scaling factor. The constant scaling factor can also be viewed as the *global normalization* of the microarray, which is used as an important pre-processing step to remove systematic bias between arrays prior to inferring differential gene expression [46, 48]. Unlike exponential distribution, scale-invariance of power-law distributions ensures non-negligible probability of occurrence at large expression values (i.e. heavy tailed). In temporal data power-law distributions are associated with presence of memory whereas exponential distributions are deemed memoryless. These differences in the statistical properties between these two classes of distributions can have far-reaching consequences on biological interpretation.

Power-law distributions as noted earlier have been observed at the gene expression level [22, 32, 44, 56]. In the present study, we investigate the validity of exponential and power-law distributional approximations at the probe PM as well as MM probe intensity levels using three different criteria, namely: $R^2$, Akaike Information Criterion (AIC) and Schwarz information criterion (SIC) [4, 7, 20, 23]. The term *approximation* is deliberately used to accommodate outliers, saturated intensities and finite sample effects inherent in microarray data. A more rigorous analysis using maximum-likelihood approach [10] may provide further insight into the distributional signatures. It is important to note that model(s) with highest $R^2$ is preferred whereas model(s) with lowest AIC and SIC are preferred. Using a combination of model validation criteria minimizes spurious conclusion that is an outcome of inherent assumptions behind a single validation criterion.

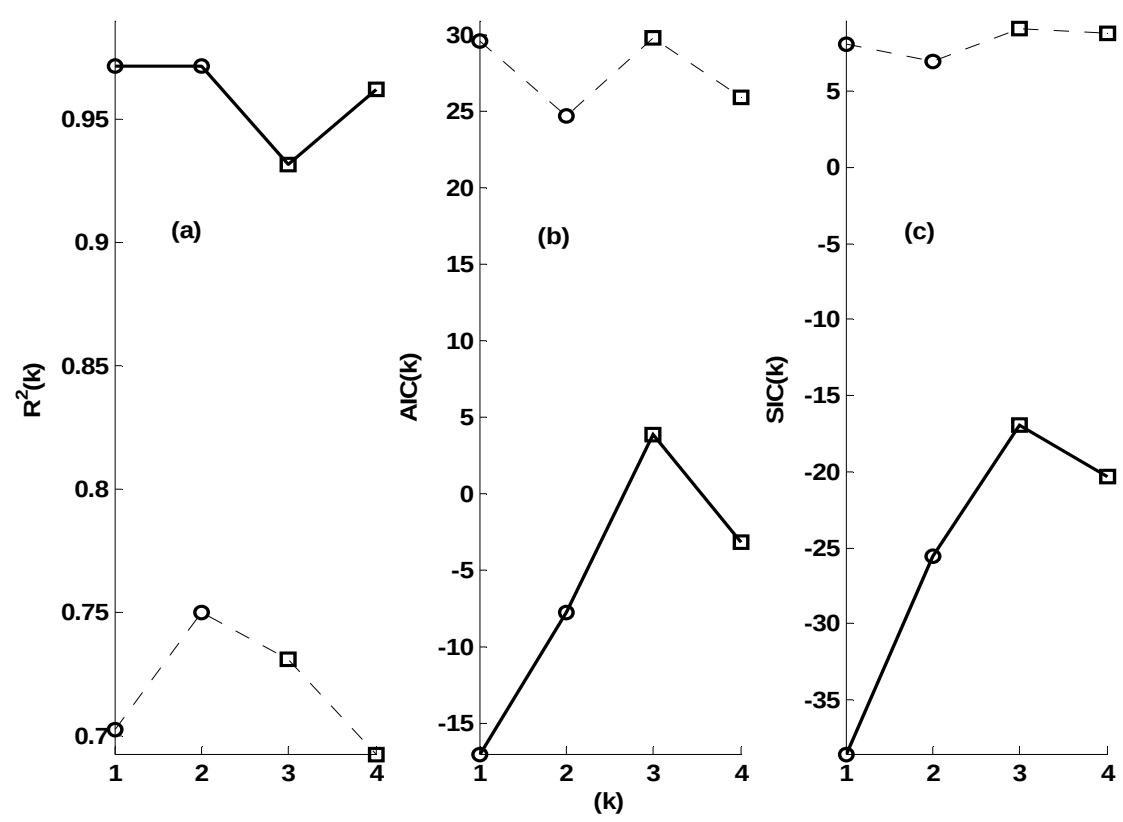

**Fig. 2.** Validation metrics ($R^2$, AIC and BIC) for the exponential (dotted lines) and power-law approximations (solid lines) across the four different distributions ($k$ = 1, 2, 3 and 4). The background subtracted gene expression data from two distinct laboratories ($L_1$, $L_2$) investigating the same paradigm [26] are represented by circles and squares respectively. While ($k$ = 1, 3) correspond to MAS 5.0, ($k$ =2, 4) correspond to RMA. The results across the three validation metrics argue in favor of power-law approximations over exponential approximations across laboratories.

Prior to model validation the distributions were log-transformed as follows:

(*i*) Transforming the exponential distribution $P(k) = \alpha_e e^{-\gamma_e k}$ yields

$$\log_2[P(k)] = \log_2(\alpha_e) - \gamma_e k, k > 0$$

(*ii*)Transforming the power-law distribution $P(k) = \alpha_p k^{-\gamma_p}$

$$\log_2[P(k)] = \log_2(\alpha_p) - \gamma_p \log_2(k), k > 0$$

Preliminary inspection of the log-log (power-law) and semi-log (exponential) plots at the probe intensity and gene expression levels revealed significant distortions for values greater than ($2^{13}$). Given the dynamic range (0, $2^{16}$-1) [1, 35, 36] of the probe intensities, it is likely that values greater than ($2^{13}$) may have significant contributions from saturated pixels. Therefore, gene expression and probe intensities above (> $2^{13}$) were filtered prior to model validation. The exponential and power-law approximations were validated using the three different criteria ($R^2$, AIC and BIC) on the filtered and background subtracted gene expression data generated across two different laboratories investigating the same paradigm generated in a recent study [26], Fig. 2. Background subtraction was accomplished by MAS 5.0 [3] and RMA [24] represented by ($k$ = 1, 3) and ($k$ = 2, 4) in Fig.2. The two different laboratories are represented by (circles, $k$ = 1, 2,) and (squares, $k$ = 3, 4) in Fig. 2, respectively. The $R^2$ values corresponding to the power-law approximation was relatively higher than that of the exponential approximation, Fig. 2a. The AIC and the BIC estimates were relatively lower for the power-law as opposed to exponential. These findings were consistent across arrays between laboratories and across background subtraction techniques. Thus power-law approximations seem to better explain the gene expression distribution as opposed to exponential approximation. These results conform to earlier findings [22, 32, 56].

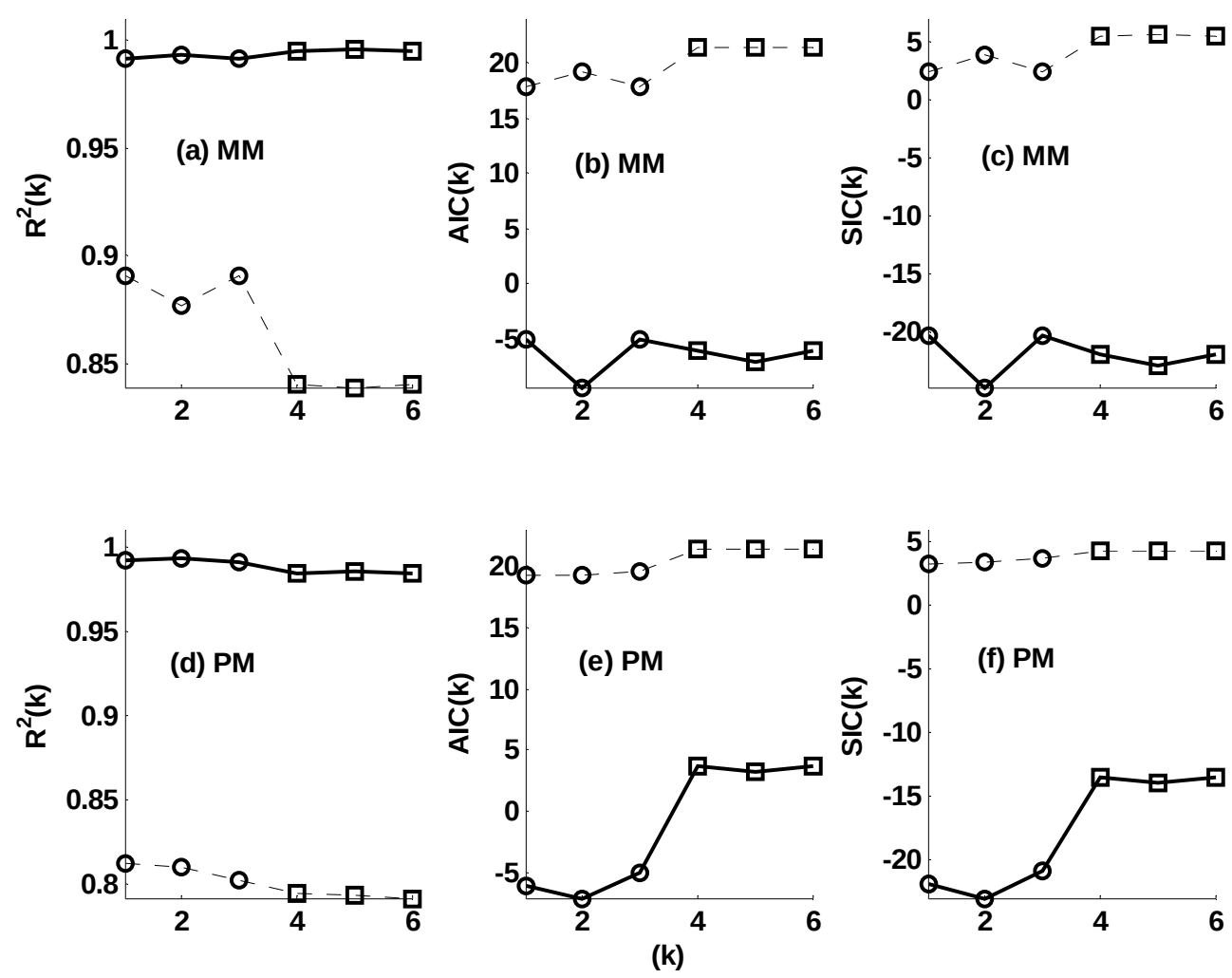

**Fig. 3.** Validation metrics ($R^2$, AIC and BIC) for the exponential (dotted lines) and power-law approximations (solid lines) across the raw and background subtracted $\pi^{MM}$ (a, b, c) and $\pi^{PM}$ (d, e, f) probe intensity distributions ($k$ = 1…6) obtained across two laboratories $L_1$,(circles) and $L_2$(squares) investigating the same paradigm [26]. The x-labels ($k$ = 1 and 4) correspond to the raw PM and MM intensities across ($L_1$, $L_2$); (k = 2 and 5) correspond to the background subtracted (MAS 5.0) PM and MM intensities across ($L_1$, $L_2$); (k = 3 and 6) correspond to background subtracted (RMA) PM and MM intensities across ($L_1$, $L_2$) respectively. The results across the three validation metrics argue in favor of power-law approximations over exponential approximations across PM as well as MM intensity distributions.

A similar analysis was carried out for the raw and background subtracted $\pi^{PM}$ and $\pi^{MM}$ probe intensities obtained from the same arrays across the same laboratories [26], Fig. 3. The raw PM and MM intensities across laboratories ($L_1$, $L_2$) are represented by ($k$ = 1 and 4), those obtained by background subtraction with MAS 5.0 and RMA are represented by ($k$ = 2 and 5) and ($k$ = 3 and 6) respectively, Fig. 3. The results obtained across the three validation criteria were consistent and argued in favor of power-law approximation over exponential approximations at the probe intensity levels.

***Remark 4*** *Power-law and exponential approximations exhibit significant difference in their statistical properties.*

- *(i)* *Analysis of the gene expression estimates across laboratories investigating the same paradigm using three validation criteria argued in favor of power-law over exponential approximations.*
- *(ii)* *Analysis of the raw and background subtracted PM and MM probe intensities in arrays across laboratories investigating the same paradigm using three validation criteria argued in favor of power-law over exponential approximations. These qualitative similarities in the distributional properties across the PM as well as MM intensities is especially intriguing as the former is a measure of specific binding whereas the latter is a measure of non-specific binding. The persistence of power-law approximations across PM and MM intensities argue in favor of non-biological factors such as static nonlinear measurement function contributing the distributional signatures.*
- *(iii)* *Power-law distributions observed at the probe intensity levels may also imply inherent clustering/patchiness in the intensities [49].*

## 3. Patchiness in PM and MM probe intensity matrices

Classical linear correlation coefficient is widely used for inferring statistically significant *linear* dependencies between a given pair of variables. Correlation coefficient between the raw and background subtracted (RMA) $\pi^{PM}$ and $\pi^{MM}$ intensities across laboratories $L_1$ (Figs. 4a and 4b) and $L_2$ (Figs. 4c and 4d) were ($r^2 \sim 0.46$, p-value < 0.05) and ($r^2 \sim 0.47$, p-value < 0.05).respectively However, visual inspection of the scatter plots, Fig. 4 revealed considerable noisiness with no apparent linear trend. Thus direct estimation of the correlation coefficient may not provide sufficient insight into their qualitative similarities and correlation structure.

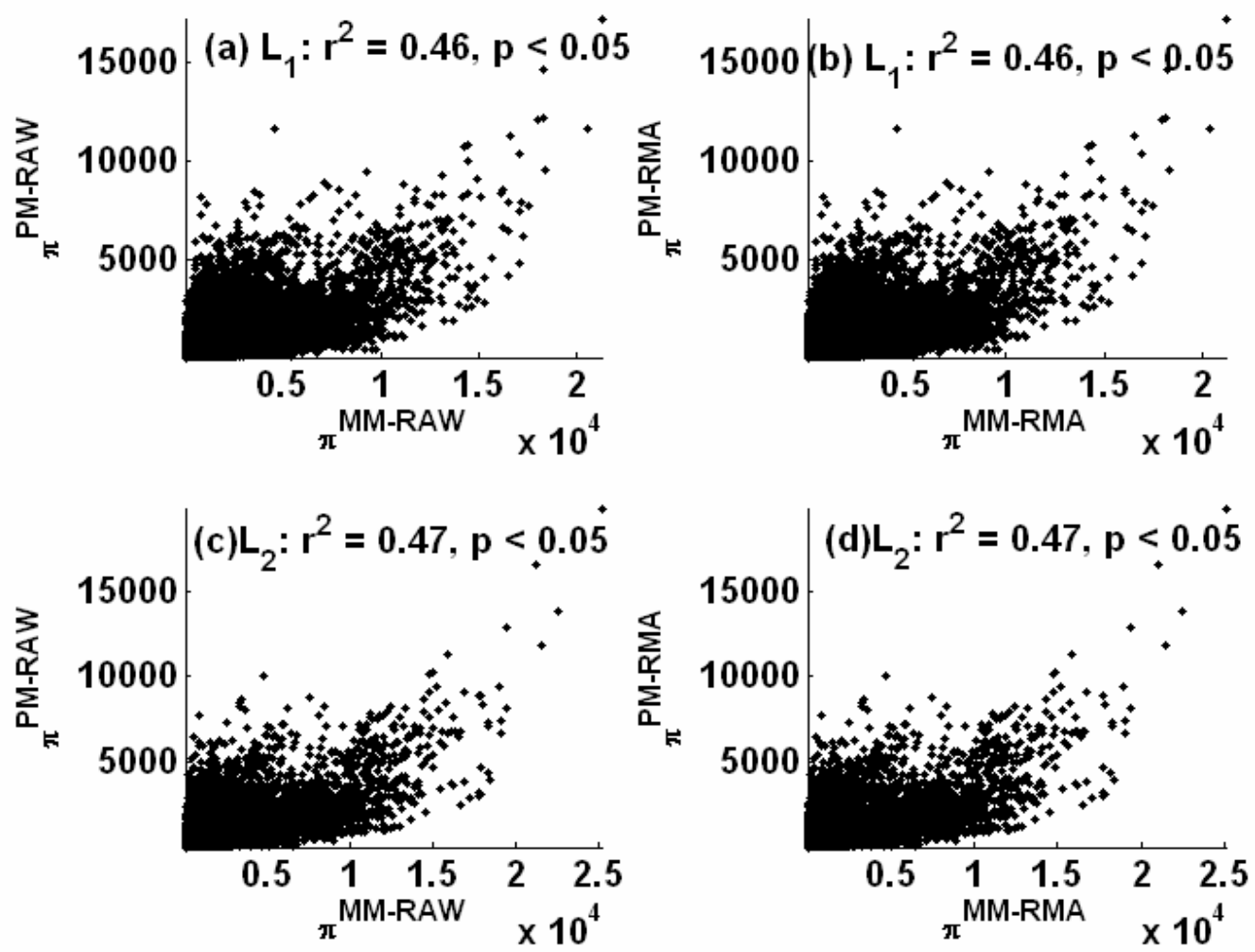

**Fig. 4.** Scatter plot of the raw and background subtracted (RMA) $\pi^{PM}$ and $\pi^{MM}$ probe intensities in arrays generated across laboratories $L_1$ (a, b) and $L_2$ (c, d) investigating the same paradigm [26].

Techniques such as *global singular-value decomposition* (SVD) [19] have been used widely in interpreting microarray gene expression data [6, 59]. Global SVD of a matrix $\Gamma$ is equivalent to eigen-decomposition of symmetric and $\Gamma^T\Gamma$ and $\Gamma\Gamma^T$, hence a measure of *linear correlation* between the probe intensities. While $\Gamma^T\Gamma$ is a measure of the row-wise correlation, $\Gamma\Gamma^T$ is a measure of the column-wise correlation. However, they both yield the same eigen-spectrum, hence equivalent.

***Remark 5*** *Classical correlation coefficient and global SVD may be useful in establishing the non-random nature of the PM and MM probe intensity matrices. However, it is possible that only a subset of the probes on the array contribute to the observed correlation. Global assessment also does not provide insights into which probes on the array contribute significantly to the observed similarity in correlation signatures between the probe intensity matrices.*

In order to overcome some of the caveats listed under Remark 5, we chose local SVD as opposed to global SVD. The procedure to determine *statistically significant patchiness* using local SVD is described in the following section.

### 3.1 Local SVD of (PM, MM) probe intensity matrices

**Algorithm I**

**Step 1** *Partition the PM probe intensity matrix* $PM^{R1xC1}$ *into non-overlapping blocks each of size r* x *c. This maps* $PM^{R1xC1}$ *into* $B^{R2xC2}$*, such that,* $\mathrm{R2} = \lfloor \mathrm{R1/r} \rfloor$*,* $\mathrm{C2} = \lfloor \mathrm{C1/c} \rfloor$ *where* $\lfloor \mathrm{y} \rfloor$ *stands for largest positive integer greater than or equal to y.*

**Step 2** *Choose a block* ***B*** *=* $B_{UV}$*, U= 1…R2, V = 1…C2. Retrieve the eigen-spectra* $\lambda_K$*, K = 1…min(R2,C2). Subsequently, normalize the eigen-values to obtain* $\delta^i = \frac{\lambda_i^2}{\sum_{i=1}^{K} \lambda_i^2} \; i = 1...K$ .

**Step 3** *Complexity* $(\eta^B)$ *of block* ***B*** *is given by*

$$\eta^B = -\frac{1}{\log K} \sum_{k=1}^{K} \delta^k \log(\delta^k)$$

*Complexity* $\eta^B$ *is inversely proportional to the linear correlation in* ***B****. Alternatively, increased redundancy/local correlation between neighboring probes in the block results in low complexity. Ideally, for a random structure the eigen-values will be uniformly distributed resulting in maximum complexity.*

**Step 4** *Block* ***B*** *is deemed as significantly correlated if the estimate of the covariance complexity on* ***B*** *is significantly different from those obtained on its random shuffled counterparts* $B_i^*, i = 1...n_s$ *of* ***B****. Random shuffled counterparts/matrices were constructed by bootstrapping the elements of* ***B*** *randomly without replacement [54, 55]. Such constrained realizations retain the distribution of the probe intensities in* ***B*** *in the shuffled counterparts whereas the spatial information between neighboring probes is destroyed.*

**Step 5** *In the presence of correlations, we expect the complexity of block* ***B*** *(*$\eta^B$*) to be lesser than that of its random shuffled counterparts* $(\eta_i^{B*}, i = 1...n_s)$*. Therefore, a one-side non-parametric test is sufficient to establish statistical significance. i.e. the null hypothesis that the given block is not significantly correlated can be rejected at a significance level* $\alpha$ *= 1/(1+*$n_s$*) if* $\eta^B < \eta_i^{B*} \forall i = 1...n_s$ *[54, 55]. In the present study,* *we fix ($n_s$ = 99), which corresponds to $\alpha$ = 0.01 [54, 55]. Parametric approaches [54, 55] are less stringent. However, their conclusions implicitly rely on implicit normality assumptions; hence can give rise to false positives when these assumptions are violated.*

**Step 6** *For visualization a binary mask* $\boldsymbol{\Phi}$ *is generated such that*

$\Phi_{UV}$ = **1** *for a significantly correlated block U= 1…R2, V = 1…C2.*

= **0** *otherwise*

*Repeat steps 2 to 5 for each of the block* $\mathbf{B} = B_{UV}$*, U= 1…R2, V = 1…C2 of the PM matrix.*

**Step 7** *Repeats Steps 1-6 independently for the (MM) probe intensity matrix.*

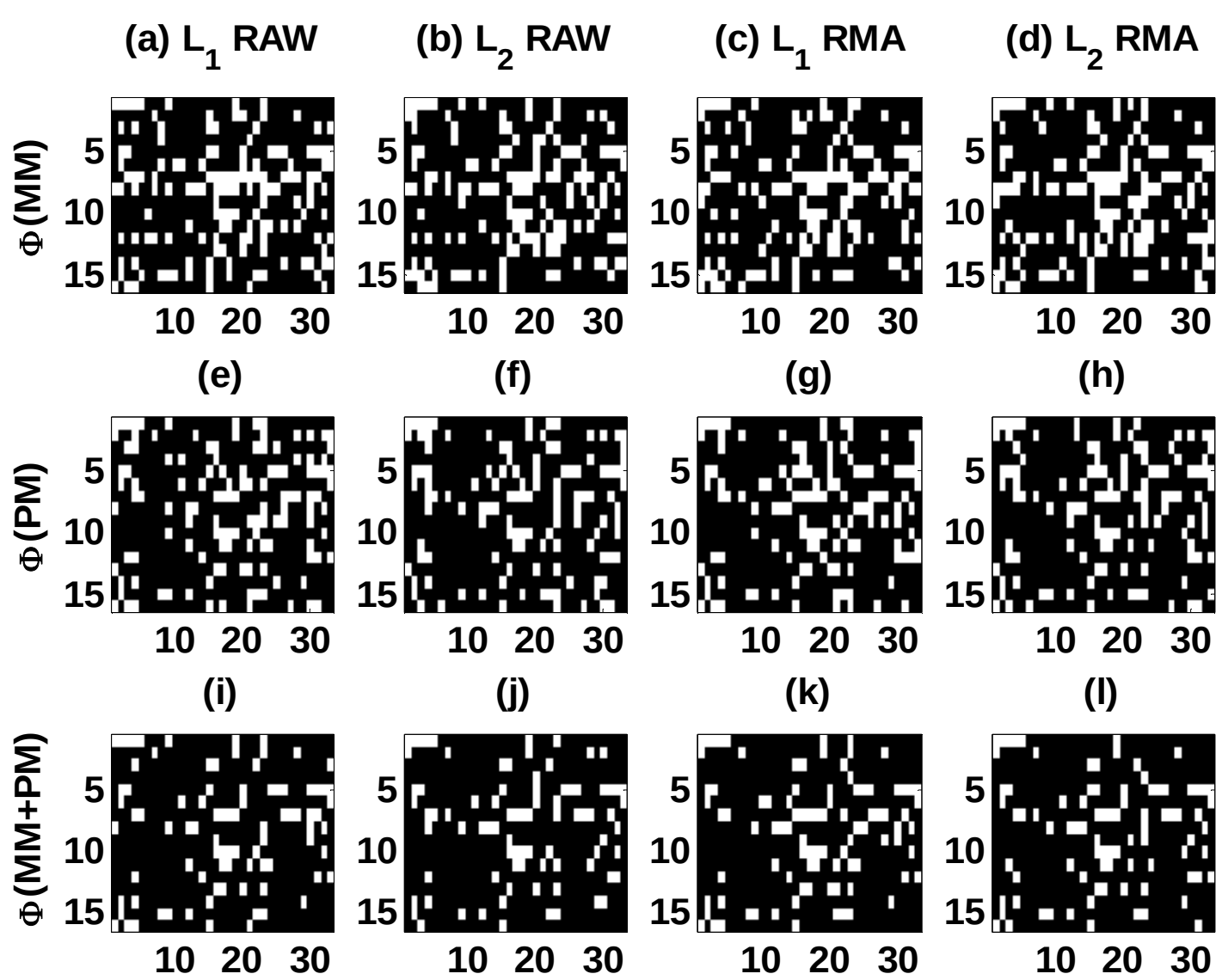

**Fig. 5.** Binary masks generated (Step 6, Algorithm I) with (r x c = 21 x 21, $n_s$ = 99) across the raw (RAW) and background subtracted (RMA) PM and MM probe intensity matrices across laboratories ($L_1$, $L_2$) investigating the same paradigm [26]. Correlated patches (white pixels) across MM, i.e. Φ(MM), and PM, i.e. Φ(PM), probe intensity matrices are shown in the top two rows (a-d and e-h), whereas those common to PM as well as MM, i.e. Φ(PM+MM), are shown in the bottom row (i-l). The size of the probe intensity matrices are (356 x 712), hence the dimension of the binary masks are (356/21 x 712/21), i.e. (16 x 33).

Global SVD is special case obtained by setting ($r$ = 1, $c$ = 1) in Step 1 of Algorithm I. As expected, complexity ($\eta$) obtained from global SVD of the PM and MM matrices with and without background subtraction were significantly lower than those of their random shuffled surrogates $\eta < \eta_i^s$, $i$ = 1… 99, indicative of non-random structure in the PM and MM matrices. This was verified across replicate arrays generated across laboratories ($L_1$, $L_2$) investigating the same given paradigm. However, from Remark 5, we note that the correlation across the probes in the PM and MM matrices *need not necessarily be global*, i.e. the statistical properties can vary considerably across the probe intensity matrices. This is to be expected as the binding efficiencies of the probes can vary considerably by their very design, also reflected by the skewed distribution of the probe intensity matrices (Sec. 2). In order to capture the local variation in correlation structure, we analyzed the probe intensity matrices using local SVD with block size (r x c = 21 x 21) and the number of surrogates ($n_s$ = 99), Fig. 5. It is important to note that there are several significantly correlated patches that persists across PM as well as MM probe intensity matrices. This is especially interesting as the former is a measure of specific binding whereas the latter is a measure of non-specific binding.

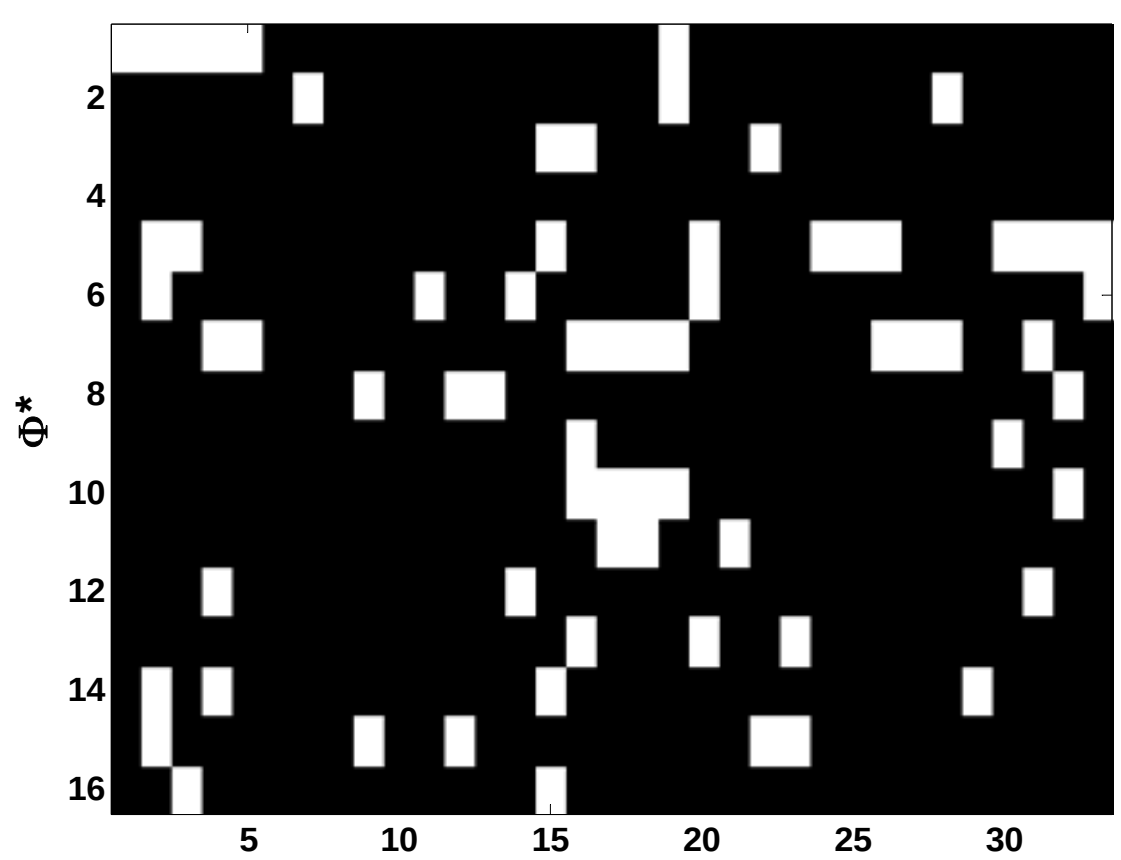

**Fig. 6.** Binary mask ($\Phi^*$) generated by intersection of the binary masks in the last row of Fig. 5, i.e.5i- 5l. The correlated patches (white pixels) in the above binary mask were common across PM and MM probe intensity matrices, across raw and background subtracted intensities and across replicate arrays generated across laboratories ($L_1$, $L_2$) [26]. The size of the probe intensity matrices are (356 x 712), hence the dimension of the masks are (356/21 x 712/21), i.e. (16 x 33).

Interestingly, there were correlated patches ($\Phi^*$) Fig. 6, that persisted (*i*) across PM and MM probe intensity matrices, (*ii*) across replicate arrays from two distinct laboratories and (*iii*) across the raw and background subtracted intensities. These patches were generated as intersection of the binary masks in Figs. 5i to 5l.

**Table 1.** Probe designations

| Suffix | Description |
|---|---|
| _f_at | Represents polymorphic probes which share considerable similarity |
| _s_at | Represents probes common across several genes/transcripts, i.e. multiple targeting |
| _g_at | Represents probes chosen in a region of overlap |
| _r_at | Represents probes picked comprising the selection rules. |
| _i_at | Represents transcripts with incomplete/fewer number of probes than required |
| _b_at | Represents ambiguous probe sets |
| _l_at | Represents transcripts with more than 20 probe pairs |
| _x_at | Represents probe-sets which share probes, i.e. non-specific binding |

The probes on the Genechip microarrays are designated based on their sequence information (see Table 1 and [3]). A recent study [42], investigated the contributions of two specific probe designations (_s_at and _x_at) on hybridization interactions and spurious correlations. Probesets with suffix (_s_at) have the ability to target multiple transcripts (i.e. multiple targeting), on the other hand those with _x_at can contribute significantly to cross-hybridization and non-specific binding. Interestingly, ~70% of the probes comprising the patchy region, Fig. 6, were classified under _s_at whereas ~11% were classified as _x_at.

***Remark 6.*** *Local SVD can be useful in identifying significantly correlated patches. Preliminary results indicated patchiness that persists across PM as well as MM probe intensity matrices with and without background subtraction. Probes that were common across the PM and MM intensity matrices, across laboratories, across raw and background subtracted data consisted mainly of cross-hybridizing and multiple targeting probes.*

*It should be noted that ($\eta$) by definition is a measure of linear correlation, hence Algorithm I can give rise to false-negatives in the presence of nonlinear correlations among the probe intensities. However, it cannot give rise to false-positives (i.e. it cannot indicate presence of correlation in a seemingly random patch). For the same reason, results obtained with ($\eta$)*

*represent the lower limit in identifying locally correlated regions. More sophisticated measures, possibly nonlinear may be used to gain further insight into the correlation structure. Algorithm I implicitly assumes a rectangular geometry, however the locally correlated regions can be irregular. This in turn may result in the inclusion/exclusions of probes which are not a member of the locally correlated region. Overlapping blocks is a suitable alternative and may be used in order to obtain finer representation of the correlation structure and minimize edge effects (i.e. accommodate all the probes on the array). The choice of block size can also affect the conclusions. A large block size provide better statistical description and especially encouraged when the probe intensity matrices are homogeneous, i.e. not much variation in the correlation properties. Small block sizes are preferred when the correlation properties show marked variations. However, smaller the block size, lesser the statistical information. There is no straightforward way to determine the optimal block size. An exhaustive approach would be to repeat Algorithm I for varying block sizes. A more elegant approach would be to use multiscale decomposition techniques such as wavelets that provide both spatial and frequency resolution.*

### 3.2 Multiscale decomposition of (PM, MM) probe intensity matrices

Multiscale approaches such as discrete wavelet transforms (DWT) are ideally suited for capturing varying statistical properties and correlation structure in 1D and 2D data. Unlike classical 2D Fourier transform (FT), DWT provides time/spatial as well as frequency resolution of the given data [37]. While high frequency components require better time resolution, low frequency components require better frequency resolution. The delicate balance between time and frequency resolutions in DWTs is dictated by the Heisenberg's uncertainty principle. DWT is a linear transform 2D FT and represents the given data as a linear combination of basis functions generated by dilating and shifting the scaling function and the mother wavelet. Dilating and shifting interrogates the correlation content in the 2D structure at various scales, hence termed as multiscale decomposition. This very aspect makes DWTs far more superior to techniques such as STFT and local SVD which captures the correlation structure at a single scale. DWT coefficients at lower-scales provide finer resolution (details) and high frequency (H) components in the given data. Those at higher-scales provide coarser resolution (approximations) or low-frequency (L) components in the given data. Since the objective of the proposed study is to un-derstand local correlation structures and their variation across the (PM, MM) probe intensity matrices, the emphasis will be on the approximation coefficients in the DWTs.

**Example** A ($k$ = 3) level hierarchical decomposition of $X$ into details and approximations using 1D DWT is shown below. The details and the approximations correspond to high-frequency (H) and low-frequency (L) components respectively. Thus at each stage one encounters two possibilities (H and L).

| | | |
|---|---|---|
| $X$ | | (given data) |
| $= L_1$ | $+ H_1$ | ($k$ = 1, first level decomposition) |
| $L_1 = L_2$ | $+ H_2$ | ($k$ = 2, second level decomposition) |
| $L_2 = L_3$ | $+ H_3$ | ($k$ = 3, third level decomposition) |

At each level ($k$), the relation $L_{k-1} = L_k + H_k$ holds. 2D DWT [37] is given as a tensor product of row-wise and column-wise 1D (separable) DWTs of the given matrix. Row-wise and column-wise decompositions give rise to approximations and details along either directions resulting in four possible outcomes namely: (LL, LH, HL and HH) respectively. Similar to 1D DWT, 2D DWT decomposition at the level $k$ satisfies the relation $LL_{k-1} = LL_k + LH_k + HL_k + HH_k$. The term $LL_k$ corresponds to the approximation (low frequency component) whereas ($LH_k$, $HL_k$ and $HH_k$) correspond to vertical, horizontal and diagonal details (high frequency components) respectively. The choice of a particular wavelet is dictated by important properties. These include (a) compact support (b) symmetry (c) orthogonality (d) regularity and (e) vanishing moments. A brief explanation of these terms are enclosed below. (a) *compact support*: wavelets with compact support correspond to FIR (finite impulse response) filters and useful in time localization. (b) *symmetry*: symmetric wavelets do not give rise to artifacts at the boundaries (c) *orthogonality*: orthogonality significantly reduces the computational burden, hence results in faster implementation (d) *regularity*: governs the degree of smoothness and usually proportional to the order of the filters. (e) *vanishing moments*: the maximum polynomial degree representation that can be generated by the scaling function. From the perspective of the proposed study, emphasis will be on (a) compact support, (b) symmetry (d) regularity and (e) vanishing moments. As noted earlier, DWT represents the correlation structure in the given matrix as a hierarchical decomposition that satisfies the recursive relation $LL_{k-1} = LL_k + LH_k + HL_k + HH_k$ at each level $k$.

A three level hierarchical decomposition (DWT) of a portion of the raw MM and the corresponding PM probe intensity matrices using Biorthogonal wavelet 2.6 (i.e. order of reconstruction = 2 and order of decomposition = 6) is shown in Figs. 7a and 7b respectively. The choice of biorthogonal wavelet is encouraged by the fact that it is compact and symmetric. The approximations at the three levels are represented by $LL_i$, $i$ = 1, 2 and 3, the horizontal, vertical and diagonal details are represented by $LH_i$, $HL_i$ and $HH_i$ respectively with $i$ = 1, 2 and 3. The magnitudes of the coefficients are color coded to aid visualization of locally correlated regions. The corresponding color-coefficient mapping is also included. Brighter colors correspond to probes which exhibit significant local correlation/patchy regions. From Figs. 4a and 4b, there is a clear overlap in local correlation structures between the PM and MM probe intensity matrices. In the following section, we propose an approach to determine whether the correlation structures are statistically significant.

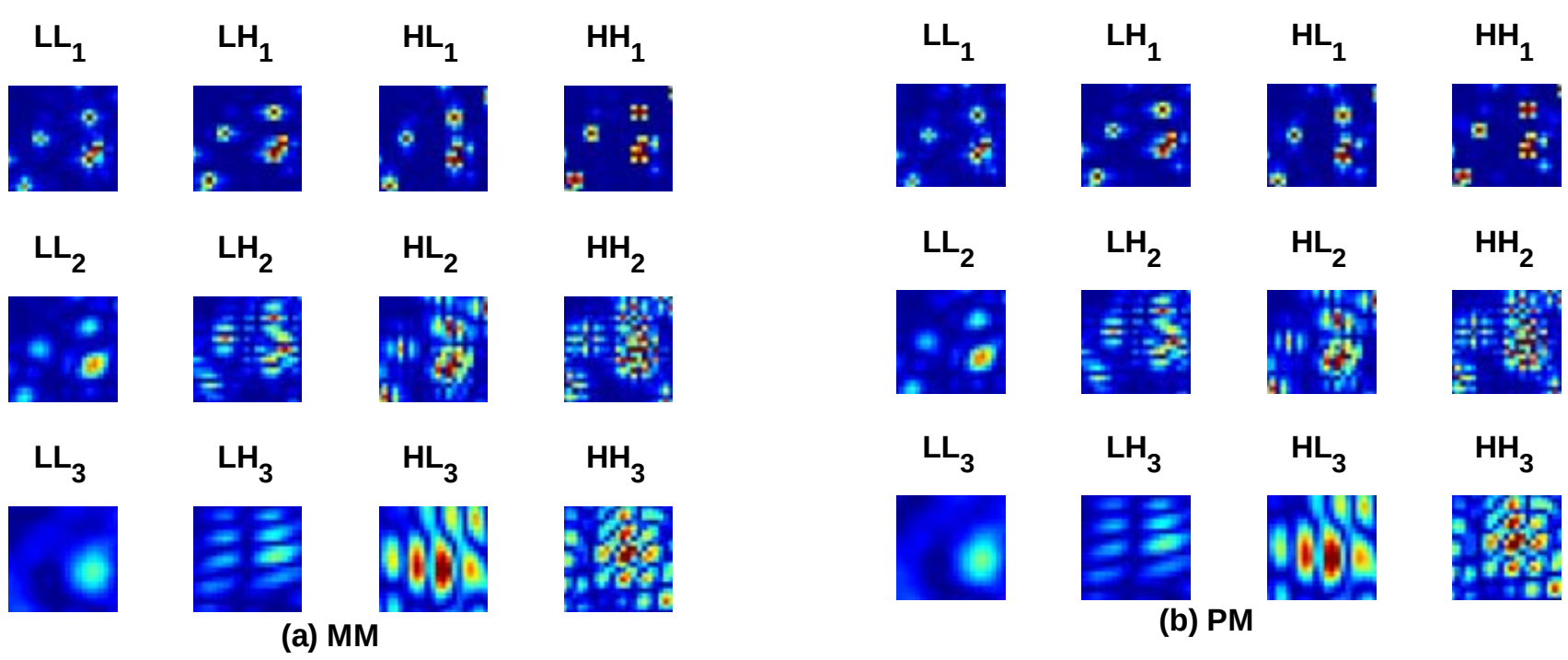

**Fig. 7.** Three-level hierarchical decomposition (DWT) of a small portion of the raw MM probe intensity matrix (a) and raw PM intensity matrix (b) using biorthogonal wavelet 2.6. The details and the approximation coefficients are color coded for visualization.

## 4. Discussion

Gene expression estimation in Genechip microarrays are governed by the qualitative behavior of atomic entities on the arrays called probes. These probes can be broadly classified into perfect and mismatch probes. While the former is a measure of specific binding, the latter is used an in-ternal control to assess non-specific binding. Understanding the qualitative behavior at the probe level can have significant impact on gene expression estimation, higher level analyses and subsequent biological inference. Classical techniques estimate gene expression as a complex combination of PM or PM and MM intensities. The behavior of the mismatch probes has especially proven to be elusive. The present study elucidates qualitative similarities in the distributional signatures and local correlation structure/patchiness of the perfect match and mismatch probe intensity matrices. The results were established on publicly available microarray gene expression data generated across laboratories investigating the same biological paradigm. These results were also established on the raw and background subtracted PM and MM probe intensity data. Thus background subtraction using popular techniques seem to have negligible effect on the qualitative similarities between PM and MM probe intensities.

Power-law approximations attributed to inherent biological mechanisms were found to persist across the PM as well as MM probe intensities and across replicate arrays generated across laboratories investigating the same paradigm. These preliminary findings argue in favor of non-biological factors contributing to the observed power-law signatures including the transfer function of the measurement device (i.e. microarray) which maps the true biological phenomena onto the probe intensity value. Analysis of the PM and MM probe intensity matrices using local singular value decomposition revealed statistically significant locally correlated patches reflecting inherent heterogeneity and variation in statistical properties. Patchiness persisted across the PM and MM probe intensity matrices. The results were established across the raw as well as background subtracted probe intensity data and across replicate arrays between laboratories investigating the same paradigm. Majority of the probes comprising the patchy regions were found to be either multiple targeting or cross-hybridizing probes. The preliminary results reported in this study raise fundamental concerns in interpreting gene expression data and encourage possible exclusion of certain probes that are common to PM as well as MM probe intensity matrices from gene expression estimation and subsequent higher level analysis. A more detailed investigation using sophisticated approaches such as maximum likelihood and multiscale decomposition is necessary in order to completely understand the distributional signatures and local correlation structures at the probe intensities.